\documentclass[12pt,onecolumn,english,technote,onecolumn]{IEEEtran}
\usepackage{amsmath}
\usepackage{newtxmath}
\usepackage[latin9]{inputenc}
\usepackage{color}
\usepackage{babel}
\usepackage{float}
\usepackage{graphicx}
\usepackage{setspace}
\usepackage{esint}
\usepackage[pdfusetitle,
 bookmarks=true,bookmarksnumbered=false,bookmarksopen=false,
 breaklinks=false,pdfborder={0 0 1},backref=false,colorlinks=true]
 {hyperref}
\hypersetup{
 pdfborderstyle=,linkcolor=blue,citecolor=blue,urlcolor=blue}

\makeatletter

\providecommand{\tabularnewline}{\\}

\usepackage[font=normalsize]{caption}
\usepackage[noadjust]{cite}

\usepackage{varwidth}

\makeatother

\begin{document}
\title{Acceleration of Convergence of Double Series for the Green's Function
of the Helmholtz Equation in Polar Coordinates}
\author{Igor M. Braver}
\maketitle
\begin{abstract}
The well-known expressions for the Green\textquoteright s functions
for the Helmholtz equation in polar coordinates with Dirichlet and
Neumann boundary conditions are transformed. The slowly converging
double series describing these Green's functions are reduced by means
of successive subtraction of several auxiliary functions to series
that converge much more rapidly. A method is given for constructing
the auxiliary functions (the first one is identical with the Green\textquoteright s
function of the Laplace equation) in the form of single series and
in the form of closed expressions. Formulas are presented for summing
series of the Fourier--Bessel and Dini type; they are required to
implement the two steps of the procedure for accelerating convergence.

The effectiveness of the functions constructed is illustrated by the
numerical solution of the problem of the dispersion properties of
a slotted line with a coaxial circular cylindrical screen.
\end{abstract}

\section*{INTRODUCTION}

\thispagestyle{empty}Let us consider Green\textquoteright s function

\begin{equation}
G_{p}(\rho,\varphi;\rho_{1},\varphi_{1})=-\frac{\delta_{ph}}{\pi\varkappa^{2}}+\sum_{\begin{subarray}{c}
m\,=\,0\\
n\,=\,1
\end{subarray}}^{\infty}\frac{\varepsilon_{m}C_{pmn}}{\pi\left(\varkappa_{pmn}^{2}-\varkappa^{2}\right)}J_{m}\left(\varkappa_{pmn}\rho\right)J_{m}\left(\varkappa_{pmn}\rho_{1}\right)\cos m\left(\varphi-\varphi_{1}\right),
\end{equation}
which, inside the unit circle, satisfies the inhomogeneous Helmholtz
equation

\begin{equation}
\Delta G_{p}+\varkappa^{2}G_{p}=-\delta\left(\mathbf{r}-\mathbf{r}_{1}\right)
\end{equation}
and the homogeneous Dirichlet ($p=e$) or Neumann ($p=h$) boundary
conditions on the circle $\rho=1$. Here $\rho$ and $\varphi$ are
polar coordinates; $\varkappa_{pmn}$ are the positive roots of the
equation $J_{m}\left(\varkappa_{emn}\right)=0$ or $J_{m}^{\,\prime}\left(\varkappa_{hmn}\right)=0$,
$J_{m}$ and $\mathit{J_{m}^{\,\prime}}$ denote the Bessel function
and its derivative; $\varepsilon_{m}=2-\delta_{m0}$, where $\delta_{mn}$
is the Kronecker delta;
\begin{equation}
C_{emn}=J_{m\,+\,1}^{-2}\left(\varkappa_{emn}\right),\qquad C_{hmn}=\frac{\varkappa_{hmn}^{2}}{\varkappa_{hmn}^{2}-m^{2}}\,J_{m}^{-2}\left(\varkappa_{hmn}\right);
\end{equation}
$\Delta$ is the two-dimensional Laplace operator, $\delta(...)$
is the Dirac function, and $\mathbf{r}$ is the radius vector of a
point inside the unit circle. Green's function (1) is employed to
formulate different integral equations in mathematical physics, in
particular, to solve electrodynamics problems concerning the characteristic
modes of a circular waveguide with infinitely thin ideal conductors
{[}1{]} or resistive films {[}2{]}. In the solution of such integral
equation (for example, by Galerkin method) it is necessary to calculate
repeatedly the double series arising as a result of the integration
of (1) with some basis functions. To carry out the calculations with
acceptable accuracy, special methods for accelerating the convergence
of such series must be employed, even when using modern computers.

\pagebreak{}

\noindent It is well-known that the double series in (1) can be reduced
to a single series using the equality {[}3{]}

\begin{equation}
\sum_{n\,=\,1}^{\infty}\frac{C_{pmn}}{\varkappa_{pmn}^{2}-\varkappa^{2}}J_{m}\left(\varkappa_{pmn}\rho\right)J_{m}\left(\varkappa_{pmn}\rho_{1}\right)=\frac{\delta_{ph}\delta_{m0}}{\varkappa^{2}}+\frac{\pi}{4}J_{m}\left(\varkappa\rho_{<}\right)\left[J_{m}\left(\varkappa\rho_{>}\right)X_{pm}(\varkappa)-Y_{m}\left(\varkappa\rho_{>}\right)\right],
\end{equation}
where $\rho_{<}=\min\left\{ \rho,\rho_{1}\right\} $, $\rho_{>}=\max\left\{ \rho,\rho_{1}\right\} $
; $X_{em}(\varkappa)=\dfrac{Y_{m}(\varkappa)}{J_{m}(\varkappa)}$,
$X_{hm}(\varkappa)=\dfrac{Y_{m}^{\,\prime}(\varkappa)}{J_{m}^{\,\prime}(\varkappa)}$,
and $Y_{m}$ and $Y_{m}^{\,\prime}$ are the Neumann function and
its derivative.

This transformation considerably simplifies the calculations if the
contour of integration lies on the circle $\rho={\rm const}$. In
so doing, integrals of the product of the basis functions and $\cos m\left(\varphi-\varphi_{1}\right)$
appear, which makes it possible to specify not the basis functions
themselves, but rather their Fourier transforms {[}4{]}. In this problem,
we arrive at single series, the general term of which is expressed
in terms of Bessel and Neumann functions, and no fundamental difficulties
are encountered in further calculations. If, however, the coordinate
$\rho$ varies along the contour of integration, it is not desirable
to reduce (1) to a single series, since it becomes necessary to carry
out a numerical integration of basis functions multiplied by the right-hand
side of (4), which depends on the sought for parameter $\varkappa$.
For this reason, in what follows, we shall examine a possible method
for accelerating the convergence of the double series (1) directly.

\section*{DERIVATION OF THE SUMMATION FORMULAS}

It is easy to verify that expression (1) can be reduced to the form

\begin{equation}
G_{p}=-\frac{\delta_{ph}}{\pi\varkappa^{2}}+G_{p1}+\varkappa^{2}G_{p2}+\cdots+\varkappa^{2j-2}G_{pj}+\varkappa^{2j}\sum_{\begin{subarray}{c}
m\,=\,0\\
n\,=\,1
\end{subarray}}^{\infty}\frac{\left(\cdots\right)}{\varkappa_{pmn}^{2j}}.
\end{equation}
Here the dots in parentheses denote the general term of the series
in (1),

\begin{equation}
G_{pj}=\sum_{m\,=\,0}^{\infty}\frac{\varepsilon_{m}}{\pi}S_{pmj}\left(\rho;\rho_{1}\right)\cos m\left(\varphi-\varphi_{1}\right),
\end{equation}

\begin{equation}
S_{pmj}\left(\rho;\rho_{1}\right)=\sum_{n\,=\,1}^{\infty}\frac{C_{pmn}}{\varkappa_{pmn}^{2j}}J_{m}\left(\varkappa_{pmn}\rho\right)J_{m}\left(\varkappa_{pmn}\rho_{1}\right).
\end{equation}

It should be noted that $G_{p}$ can be expanded in an infinite power
series in the parameter $\varkappa$ (i.e., the limit $j\rightarrow\infty$
can be taken in (5)) only under the additional condition that $|\varkappa$\textbar{}
is less than any root $\varkappa_{pmn}$. Since $j$ is finite and
the last term in (5) is preserved, this strict equality remains valid
for any value of $\varkappa$.

It will be shown below that simpler representations can be constructed
for $G_{pj}$, and therefore the general term in the double series
describing $G_{p}$ acquires, owing to the transformation (5), the
additional factor $\varkappa_{pmn}^{-2j}$.

\pagebreak{}

We shall now simplify the functions $G_{pj}$. The most important
part of the further transformations is to obtain closed-form expressions
for the single series $S_{pmj}$. According to the theory of Fourier-Bessel
and Dini series {[}5{]}, we have

\begin{equation}
S_{pm0}\left(\rho;\rho_{1}\right)=-\delta_{ph}\delta_{m0}+\dfrac{1}{2\rho}\delta\left(\rho-\rho_{1}\right).
\end{equation}
For $j=1,2,3,...$ the functions $S_{pmj}$ are related by the recurrence
relation

\begin{equation}
\frac{1}{\rho}\frac{\partial}{\partial\rho}\left[\rho\frac{\partial}{\partial\rho}S_{pmj}\right]-\frac{m^{2}}{\rho^{2}}S_{pmj}=-S_{pm,\,j\,-\,1}.
\end{equation}
The required expressions for $S_{pmj}$ can be constructed by solving
the inhomogeneous differential equations (9) with Dirichlet or Neumann
boundary conditions. In the special case when $p=h$ and $m=0$, equations
(9) and the Neumann conditions determine $S_{h0j}$, up to an additive
constant. To calculate this constant it is necessary to use the orthogonality
relation

\begin{equation}
\intop_{0}^{1}S_{h0j}\rho d\rho=0,
\end{equation}
which follows from the representation of $S_{h0j}$ in the form of
the series (7).

We shall present the expressions for $S_{pmj}$ obtained for $j=1$:

\begin{equation}
\left\{ \begin{aligned}S_{e01} & =-\frac{1}{2}\ln\rho_{>},\\
S_{em1} & =\frac{1}{4m}\left[\left(\frac{\rho_{<}}{\rho_{>}}\right)^{m}-\left(\rho\rho_{1}\right)^{m}\right],\:m\geq1;
\end{aligned}
\right.
\end{equation}

\begin{equation}
\left\{ \begin{aligned}S_{h01} & =\frac{1}{4}\left(\rho^{2}+\rho_{1}^{2}\right)-\frac{3}{8}-\frac{1}{2}\ln\rho_{>},\\
S_{hm1} & =\frac{1}{4m}\left[\left(\frac{\rho_{<}}{\rho_{>}}\right)^{m}+\left(\rho\rho_{1}\right)^{m}\right],\:m\geq1
\end{aligned}
\right.
\end{equation}

and for $j=2$

\begin{equation}
\left\{ \begin{aligned}S_{e02} & =\frac{1}{8}\left[\left(\rho^{2}+\rho_{1}^{2}\right)\ln\rho_{>}+1-\rho_{>}^{2}\right],\\
S_{e12} & =\frac{1}{32}\left[\rho\rho_{1}\left(\rho^{2}+\rho_{1}^{2}-1\right)-4\rho\rho_{1}\ln\rho_{>}-\frac{\rho_{<}^{3}}{\rho_{>}}\right],\\
S_{em2} & =\frac{1}{16m(m^{2}-1)}\biggl\{\left(\rho\rho_{1}\right){}^{m}\left[(\rho^{2}+\rho_{1}^{2})(m-1)-2m\right]\\
 & \quad+\left(\frac{\rho_{<}}{\rho_{>}}\right)^{m}\left[\rho_{>}^{2}(m+1)-\rho_{<}^{2}(m-1)\right]\biggr\},\quad m\geq2;
\end{aligned}
\right.
\end{equation}
\medskip{}

\begin{equation}
\left\{ \begin{aligned}S_{h02} & =\frac{7}{192}-\frac{1}{64}\left(\rho^{4}+\rho_{1}^{4}\right)+\frac{3}{32}\left(\rho^{2}+\rho_{1}^{2}\right)-\frac{1}{16}\rho^{2}\rho_{1}^{2}+\frac{1}{8}\left(\rho^{2}+\rho_{1}^{2}\right)\ln\rho_{>}-\frac{1}{8}\rho_{>}^{2},\\
S_{h12} & =\frac{1}{32}\left[\rho\rho_{1}\left(7-\rho^{2}-\rho_{1}^{2}\right)-4\rho\rho_{1}\ln\rho_{>}-\frac{\rho_{<}^{3}}{\rho_{>}}\right],\\
S_{hm2} & =\frac{1}{16m(m^{2}-1)}\biggl\{\left(\rho\rho_{1}\right){}^{m}\left[\frac{2m^{2}-4}{m}-\left(\rho^{2}+\rho_{1}^{2}\right)(m-1)\right]\\
 & \quad+\left(\frac{\rho_{<}}{\rho_{>}}\right)^{m}\left[\rho_{>}^{2}(m+1)-\rho_{<}^{2}(m-1)\right]\Biggr\},\quad m\geq2.
\end{aligned}
\right.
\end{equation}
\medskip{}

Expressions (11) and (12) corresponding to $m\geq1$ were derived
previously in {[}6{]} by a different method, based on the expansion
of the left-hand and right-hand sides of equation (4) in a power series
in the parameter $\varkappa$ and equating the coefficients in front
of $\varkappa^{0}$.

Substituting representations (11)--(14) into formula (6) we obtain
a series which can be summed over the index $m$ analytically. For
$j=1$ we obtain

\begin{equation}
G_{e1}=\frac{1}{2\pi}\ln\frac{\overline{R}}{R},\quad G_{h1}=\frac{1}{4\pi}\left(\rho^{2}+\rho_{1}^{2}\right)-\frac{3}{8\pi}-\frac{1}{2\pi}\ln\left(R\overline{R}\right);
\end{equation}

\begin{equation}
\begin{aligned}R & =\sqrt{\rho^{2}+\rho_{1}^{2}-2\rho\rho_{1}\cos\left(\varphi-\varphi_{1}\right)},\\
\overline{R} & =\sqrt{1+\rho^{2}\rho_{1}^{2}-2\rho\rho_{1}\cos\left(\varphi-\varphi_{1}\right)}.
\end{aligned}
\end{equation}

We note that $G_{p1}$ is Green's function of the Laplace equation
$\Delta G_{p1}=-\delta\left(\mathbf{r}-\mathbf{r}_{1}\right)+\dfrac{\delta_{ph}}{\pi}$,
related to Green's function of the Helmholtz equation by the relation

\begin{equation}
G_{p1}=\lim_{\varkappa\rightarrow0}\left(G_{p}+\dfrac{\delta_{ph}}{\pi\varkappa^{2}}\right).
\end{equation}

The logarithmic terms in formulas (15) can be easily found using the
method of images. The applicability of this method for regions with
different configurations is examined in detail in {[}7{]}. The new
result in (15) is only the regular term $\dfrac{\left(\rho^{2}+\rho_{1}^{2}-1.5\right)}{4\pi}$,
describing the difference between the function $G_{h1}$ and its representation
by the method of images, which in the case of the Neumann problem
for a circular region can be applied only approximately {[}7{]}. The
use of $G_{p1}$ to accelerate the convergence of the series describing
$G_{p}$ was also proposed in {[}7{]}. The transformation $G_{p}=G_{p1}+\left(G_{p}-G_{p1}\right)$
becomes inefficient when the kernel of the integral equation contains
the second derivative of $G_{p}$ along the normal to the integration
contour {[}2{]}, since the difference series $G_{p}-G_{p1}$, after
the double differentiation, converges just as slowly as the starting
series $G_{p}$. Further transformations can be based on the use of
$G_{p2}$. Expressions (6), (13), and (14) give a representation of
$G_{p2}$, in the form of single series. This representation is convenient
for integrating $G_{p2}$ with basis functions containing only integer
powers of $\rho$ {[}2{]}. The function $G_{p2}$ can be constructed
in closed form in the same way as $G_{p1}$. For example, for the
tensor $\overline{\mathbf{G}}_{st}$, introduced in {[}1{]} and related
to $G_{h2}$ by the relation

\begin{equation}
\overline{\mathbf{G}}_{st}=\left(\mathbf{a}_{\rho}\frac{\partial}{\rho\partial\varphi}-\mathbf{a}_{\varphi}\frac{\partial}{\partial\rho}\right)\left(\mathbf{a}_{\rho_{1}}\frac{\partial}{\rho_{1}\partial\varphi_{1}}-\mathbf{a}_{\varphi_{1}}\frac{\partial}{\partial\rho}\right)G_{h2},
\end{equation}
where $\mathbf{a}_{\rho}$, $\mathbf{a}_{\varphi}$ are the unit vectors
of the polar coordinates, we obtain an expression that is identical
to the one obtained in {[}1{]}.

\section*{EXAMPLE OF NUMERICAL REALIZATION}

To illustrate the effectiveness of the representations constructed
for $G_{p}$ we shall find the wave number of the dominant mode of
a screened slotted line (Fig. 1). This problem is not discussed in
the detailed reviews in {[}8, 1{]} of the literature on waveguides
with complicated shapes. Such a line has apparently not been analyzed
previously.

\begin{figure}[H]
\begin{centering}
\includegraphics{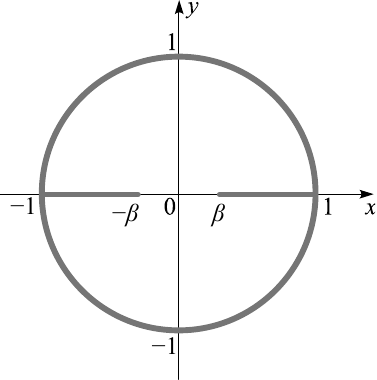}\caption{\label{fig:1}Cross section of a circular waveguide with a slotted
line in the diametral plane.}
\par\end{centering}
\end{figure}

The tangential electric field $E_{x}$ in the plane of the slot satisfies
the integral equation

\begin{equation}
\intop_{0}^{\beta}Q\left(\rho;\rho_{1}\right)E_{x}\left(\rho_{1}\right)d\rho_{1}=0,\quad0\leq\rho\leq\beta,
\end{equation}
where $Q(\rho;\rho_{1})$ is the Green's function $G_{h}$ calculated
for $\varphi=\varphi_{1}$, in which all terms with an odd index $m$
are dropped. The evenness of the field of the mode under study $E_{x}(-x)=E_{x}(x)$
was taken into account in writing (19). We will seek the solution
of the integral equation (19) in the form

\begin{equation}
E_{x}(\rho)=\sum_{\nu\,=\,0}^{N}A_{\nu}\frac{T_{2\nu}(u)}{\sqrt{1-u^{2}}},\quad u=\frac{\rho}{\beta},
\end{equation}
$A_{\nu}$ are unknown coefficients and $T_{\nu}(u)=\cos(\nu\arccos u)$
are Chebyshev polynomials. Implementation of Galerkin method leads
to the dispersion equation

\begin{equation}
\det\left\Vert Z_{\mu\nu}\right\Vert =0;\quad\mu,\nu=0,1,\ldots,N,
\end{equation}
with matrix elements

\begin{equation}
Z_{\mu\nu}=-\frac{1}{\varkappa^{2}}\delta_{\mu0}\delta_{\nu0}+\sigma_{\mu\nu1}+\varkappa^{2}\sigma_{\mu\nu2}+\varkappa^{4}\sum_{\begin{subarray}{c}
m=0,2,4,\ldots\\
n=1,2,3,\ldots
\end{subarray}}\frac{\varepsilon_{m}C_{hmn}}{\varkappa_{hmn}^{2}-\varkappa^{2}}\frac{I_{\mu mn}I_{\nu mn}}{\varkappa_{hmn}^{4}}.
\end{equation}
Here the values

\begin{equation}
I_{\nu mn}=J_{\frac{m}{2}+\nu}\left(\frac{1}{2}\varkappa_{hmn}\beta\right)J_{\frac{m}{2}-\nu}\left(\frac{1}{2}\varkappa_{hmn}\beta\right)
\end{equation}
were obtained from formula (7.361) of {[}9{]}, while the quantities
$\sigma_{\mu\nu1},\,\sigma_{\mu\nu2}$ are obtained by integrating
$G_{h1},\,G_{h2}$ with basis function (20) and have the following
form:

\begin{subequations}
\begin{equation}
\begin{aligned}\sigma_{\mu\nu1} & =\left(\dfrac{\beta^{2}}{4}-\dfrac{3}{8}-\dfrac{1}{2}\ln\dfrac{\beta}{2}\right)\delta_{\mu0}\delta_{\nu0}+\dfrac{\beta^{2}}{16}\left(\delta_{\mu0}\delta_{\nu1}+\delta_{\nu0}\delta_{\mu1}\right)\\
 & +\dfrac{1}{8\mu}\left(1-\delta_{\mu0}\right)\delta_{\mu\nu}+\dfrac{1}{4}\sum_{k\,=\,1}^{\infty}\frac{1}{k}\beta^{4k}\Gamma_{k\mu}\Gamma_{k\nu},
\end{aligned}
\end{equation}
\begin{equation}
\begin{aligned}\sigma_{\mu\nu2} & =\frac{7}{192}\delta_{\mu0}\delta_{\nu0}-\frac{\beta^{4}}{64}\left(\Gamma_{2\mu}\delta_{\nu0}+\Gamma_{2\nu}\delta_{\mu0}\right)-\dfrac{\beta^{4}}{16}\Gamma_{1\mu}\Gamma_{1\nu}-\dfrac{\beta^{2}}{32\mu}\left(1-\delta_{\mu0}\right)\delta_{\mu\nu}\\
 & -\dfrac{\beta^{2}}{128\mu}\left(1-\delta_{\mu0}\right)\left(\delta_{\mu,\nu+1}+\delta_{\mu,\left|\nu-1\right|}\right)-\dfrac{\beta^{2}}{128\nu}\left(1-\delta_{\nu0}\right)\left(\delta_{\mu+1,\nu}+\delta_{\left|\mu-1\right|,\nu}\right)\\
 & +\dfrac{\beta^{2}}{32}\left(\frac{1}{2\mu+1}\delta_{\mu\nu}+\frac{1}{2\mu+1}\delta_{2\mu+1,\left|2\nu-1\right|}+\frac{1}{2\nu+1}\delta_{\left|2\mu-1\right|,2\nu+1}+\frac{1}{\left|2\mu-1\right|}\delta_{\left|2\mu-1\right|,\left|2\nu-1\right|}\right)\\
 & +\dfrac{\beta^{2}}{32}\left(4\ln\frac{\beta}{2}-1\right)\left(\Gamma_{1\mu}\delta_{\nu0}+\Gamma_{1\nu}\delta_{\mu0}\right)+\frac{1}{8}\sum_{k\,=\,1}^{\infty}\frac{2k^{2}-1}{k^{2}(4k^{2}-1)}\beta^{4k}\Gamma_{k\mu}\Gamma_{k\nu}\\
 & -\dfrac{\beta^{2}}{16}\sum_{k\,=\,1}^{\infty}\frac{1}{k(2k+1)}\beta^{4k}\left(\Gamma_{k+1,\mu}\Gamma_{k\nu}+\Gamma_{k\mu}\Gamma_{k+1,\nu}\right),
\end{aligned}
\end{equation}
\end{subequations}

\begin{equation}
\Gamma_{k\nu}=\left\{ \begin{aligned} & \frac{(2k)!}{2^{2k}(k+\nu)!(k-\nu)!}, & k\geq\nu,\\
 & 0, & k<\nu.
\end{aligned}
\right.
\end{equation}

To obtain (24) the functions $G_{h1}$ and $G_{h2}$ were written
in the form of single series in even indices $m$. The part of the
series whose general term is proportional to $\left(\dfrac{\rho_{<}}{\rho_{>}}\right)^{m}$
reduces as a result of summation to expressions containing $\ln\left|\rho\pm\rho_{1}\right|$,
and can be analytically integrated {[}10{]} with weighted Chebyshev
polynomials. The rest of the series can be integrated with the basis
function term by term, since the result of the integration, containing
the factor $\beta^{4k}$ in the general term, converges very rapidly
for $\beta<1$.

As a result of the improvement in the convergence based on formula
(5) with $j=2$, no additional computer time is required to sum the
double series in (22). Thus, to solve the dispersion equation (21)
to six significant figures in $\varkappa$ the series can be limited
to $m\leq8$, $n\leq5$ (25 terms). This makes it possible to employ
tabulated values of $\varkappa_{hmn}$ {[}11{]}. The total time for
calculating $\varkappa$ on an ES-1060 computer for any value of $\beta$
does not exceed several seconds. The results of the numerical calculation
are summarized in Table 1. 
\begin{center}
\begin{table}[H]
\caption{}

\centering{}%
\begin{tabular}{|c|c|c|c|c|c|}
\hline 
$\beta$ & $\varkappa$ & $\beta$ & $\varkappa$ & $\beta$ & $\varkappa$\tabularnewline
\hline 
\hline 
0.05 & 0.817917 & 0.40 & 1.39609 & 0.75 & 1.77598\tabularnewline
\hline 
0.10 & 0.930063 & 0.45 & 1.46510 & 0.80 & 1.80187\tabularnewline
\hline 
0.15 & 1.02034 & 0.50 & 1.53131 & 0.85 & 1.82042\tabularnewline
\hline 
0.20 & 1.10164 & 0.55 & 1.59346 & 0.90 & 1.83252\tabularnewline
\hline 
0.25 & 1.17841 & 0.60 & 1.65014 & 0.95 & 1.83915\tabularnewline
\hline 
0.30 & 1.25262 & 0.65 & 1.70001 & -- & \tabularnewline
\hline 
0.35 & 1.32514 & 0.70 & 1.74209 & -- & \tabularnewline
\hline 
\end{tabular}
\end{table}
\par\end{center}

Since every basis function in (20) takes into account the singularity
of the field near the edge {[}12{]}, a small number of basis functions
was required to stabilize the values of $\varkappa$. For $\beta=0.95$
it is sufficient to set $N=3$ in (20); for $\beta\leq0.2$ the first
($\nu=0$) basis function already yields $\varkappa$ with a relative
error not exceeding $5\times10^{-6}$ . This is explained by the fact
that as $\beta\rightarrow0$ the function $\dfrac{1}{\sqrt{1-u^{2}}}$
is the exact solution of the integral equation (19). This is easily
verified by retaining in the kernel only terms which do not vanish
as $\beta\rightarrow0$. Using (4) we find

\begin{equation}
Q\left(\rho;\rho_{1}\right)\simeq\frac{1}{4}\frac{Y_{1}(\varkappa)}{J_{1}(\varkappa)}-\frac{1}{2\pi}\left(\gamma+\ln\frac{\varkappa\beta}{2}\right)-\frac{1}{4\pi}\ln\left|u^{2}-u_{1}^{2}\right|,\quad\beta\rightarrow0,
\end{equation}
where $\gamma=0.5772...$ is Euler's constant. Expression (26) yields
for $\varkappa$ the approximate dispersion equation

\begin{equation}
\ln\frac{4}{\varkappa\beta}+\frac{\pi}{2}\frac{Y_{1}(\varkappa)}{J_{1}(\varkappa)}-\gamma=0,
\end{equation}
which holds for $\beta\ll1$ . If the gap is logarithmically narrow
$\left(\ln\dfrac{1}{\beta}\gg1\right)$, the well-known asymptotic
behavior $\varkappa\simeq\sqrt{\dfrac{2}{\ln\frac{1}{\beta}}}$ follows
from (27).

The other limiting case $\beta\rightarrow1$ is most conveniently
studied by solving the integral equation

\begin{equation}
\intop_{\beta}^{1}T\left(\rho;\rho_{1}\right)j_{x}\left(\rho_{1}\right)d\rho_{1}=0,\quad\beta\leq\rho\leq1,
\end{equation}
for the current density $j_{x}$ on perfectly conducting ridges. For
$\beta\rightarrow1$ and $\varkappa\rightarrow\varkappa_{h11}$, the
kernel of (28) can be reduced to the form

\begin{equation}
T\left(\rho;\rho_{1}\right)\simeq\frac{\varkappa_{h11}^{2}}{\left(\varkappa_{h11}^{2}-1\right)\left(\varkappa_{h11}^{2}-\varkappa^{2}\right)}-\frac{1}{8(1-\beta)^{2}}\frac{\partial^{2}}{\partial v\partial v_{1}}\ln\left|\frac{v+v_{1}}{v-v_{1}}\right|,\;v=\frac{1-\rho}{1-\beta},\quad\beta\rightarrow1.
\end{equation}
The solution of (28) and (29) is the function $\sqrt{1-v^{2}}$. Substitution
of this function into the integral equation (28) gives a dispersion
equation, whose solution for $1-\beta\ll1$ is given by
\begin{equation}
\varkappa=\varkappa_{h11}-\frac{\varkappa_{h11}}{\varkappa_{h11}^{2}-1}(1-\beta)^{2}.
\end{equation}
The results of the calculation of $\varkappa$ using the asymptotic
formulas (27) and (30) are presented in Fig.~2 (curves \textit{1}
and \textit{2}). The figure also shows for comparison the curve \textit{3}
constructed from the data given in the table of exact values of $\varkappa$
presented above in Table 1.

\begin{figure}[H]
\centering{}\includegraphics{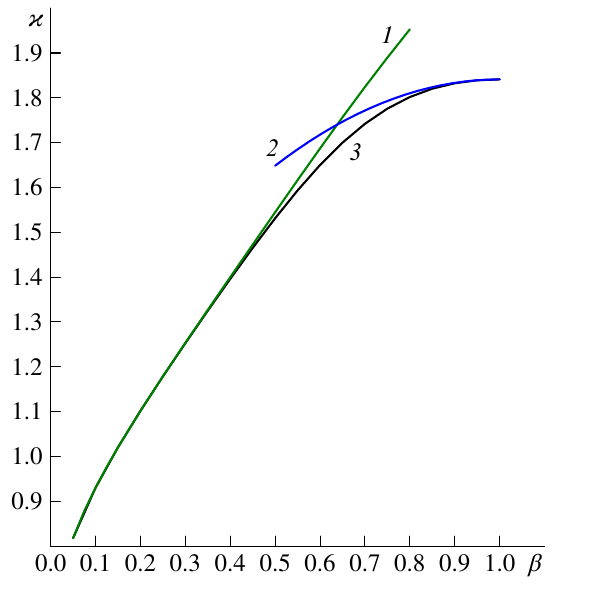}\caption{\label{fig:1-2}Wave number of the dominant mode for various widths
of the slot: \textit{1} and \textit{2} -- calculations using asymptotic
formulas (27) and (30), \textit{3} -- solution of the rigorous dispersion
equation (21).}
\end{figure}

\section*{REFERENCES}

{[}1{]} G. Conciauro, M. Bressan, and C. Zuffada, \textquotedblleft Waveguide
Modes Via an Integral Equation Leading to a Linear Matrix Eigenvalue
Problem,\textquotedblright{} \textit{IEEE Transactions on Microwave
Theory and Techniques}, vol. 32, no. 11, pp. 1495--1504, November
1984. DOI: \href{http://dx.doi.org/10.1109/TMTT.1984.1132880}{10.1109/TMTT.1984.1132880}

{[}2{]} I.\,M. Braver, Kh.\,L. Garb, and P.\,Sh. Fridberg, ``Dispersion
Equation for a Circular Waveguide with an Impedance Film,'' \textit{Soviet
Physics --- Doklady}, vol. 30, no. 1, pp. 84--86, January 1985.

{[}3{]} I.\,M. Braver, Kh.\,L. Garb, and P.\,Sh. Fridberg, ``An
Integral Equation for the Surface Current in the Problem of the Natural
Modes of a Waveguide Containing a Resistive Film,'' \textit{Soviet
Journal of Communications Technology and Electronics}, vol. 31, no.
2, pp. 33--39, February 1986.

{[}4{]} R.\,F. Fikhmanas and P.\,Sh. Fridberg, \textquotedblleft A
Method for the Numerical Application of Schwinger's Variational Principle,\textquotedblright{}
\textit{Soviet Physics --- Doklady}, vol. 21, no. 12, pp. 716--718,
December 1976.

{[}5{]} G.\,N. Watson, \textit{A Treatise on the Theory of Bessel
Functions}, Cambridge University Press, 1945.

{[}6{]} I.\,M. Braver, Kh.\,L. Garb, and P.\,Sh. Fridberg, ``Dispersion
Properties of a Circular Waveguide with a Resistive Film of Arbitrary
Width in the Diametral Plane,'' \textit{Soviet Journal of Communications
Technology and Electronics}, vol. 31, no. 4, pp. 49--54, April 1986.

{[}7{]} P.\,M. Morse and H. Feshbach, \textit{Methods of Theoretical
Physics}, McGraw-Hill, New York, 1953.

{[}8{]} F.\,L. Ng, \textquotedblleft Tabulation of Methods for the
Numerical Solution of the Hollow Waveguide Problem,\textquotedblright{}
\textit{IEEE Transactions on Microwave Theory and Techniques}, vol.
22, no. 3, pp. 322--329, March 1974. DOI: \href{http://dx.doi.org/10.1109/TMTT.1974.1128217}{10.1109/TMTT.1974.1128217}

{[}9{]} I.\,S. Gradshtein and I.\,M. Ryzhik, \textit{Table of Integrals,
Sums, Series, and Products}, Academic Press, 2007.

{[}10{]} Ch.\,M. Butler and D.\,R. Wilton, \textquotedblleft General
Analysis of Narrow Strips and Slots,\textquotedblright{} \textit{IEEE
Transactions on Antennas and Propagation}, vol. 28, no. 1, pp. 42--48,
January 1980. DOI: \href{http://dx.doi.org/10.1109/TAP.1980.1142291}{10.1109/TAP.1980.1142291}

{[}11{]} F.\,W.\,J. Olver {[}Ed.{]}, \textit{Royal Society Mathematical
Tables: Volume 7, Bessel Functions, Part 3, Zeros and Associated Values},
Cambridge University Press, 1960.

{[}12{]} J. Meixner, \textquotedblleft The Behavior of Electromagnetic
Fields at Edges,\textquotedblright{} \emph{IEEE Transactions on Antennas
and Propagation}, vol. 20, no. 4, pp. 442--446, July 1972. DOI: \href{https://dx.doi.org/10.1109/TAP.1972.1140243}{10.1109/TAP.1972.1140243}
\end{document}